\documentclass[twocolumn,amsmath,showpacs,amsfonts,aps,prc,floatfix]{revtex4}
\usepackage{graphicx}
\usepackage{bm}
\begin{document}

\title{Charged particle's  elliptic flow in 2+1D viscous hydrodynamics
 at LHC ( $\sqrt{s}$= 2.76 TeV) energy in Pb+Pb collision. }
\author{Victor Roy}
\email[E-mail:]{victor@veccal.ernet.in}
\affiliation{Variable Energy Cyclotron Centre, 1/AF, Bidhan Nagar, 
Kolkata 700~064, India}
\author{A. K. Chaudhuri}
\email[E-mail:]{akc@veccal.ernet.in}
\affiliation{Variable Energy Cyclotron Centre, 1/AF, Bidhan Nagar, 
Kolkata 700~064, India}

\begin{abstract}
In the Israel-Stewart's theory of second order hydrodynamics, we have simulated   $\sqrt{s}$=2.76 TeV Pb+Pb collisions. ALICE data for the centrality dependence of charged particles multiplicity, $p_T$ spectra in 0-5\% collisions, centrality dependence of integrated and differential elliptic flow are analysed. 
Analysis indicate that while ALICE data on charged particles multiplicity or $p_T$ spectra do not demand any viscosity, viscosity is demanded by the elliptic flow data. From a simulataneous fit to all the data sets,  
 viscosity to entropy ratio in $\sqrt{s}_{NN}$=2.76 TeV Pb+Pb collisions is extracted as, $\eta/s=0.06\pm 0.02$.
\end{abstract}

\pacs{12.38.Mh  ,47.75.+f,  25.75.Ld} 

\date{\today}  

\maketitle

\section{Introduction}

Elliptic flow is one of the most important observables in relativistic energy heavy ion collisions. Elliptic flow  measure the azimuthal correlation of produced particles with respect to the reaction plane.
Finite elliptic flow in
a non-central collision is now regarded as a definitive signature of collective effect in relativistic energy nuclear collisions \cite{Ollitrault:1992bk,Poskanzer:1998yz}. 
 Qualitatively, elliptic flow is best  explained in a relativistic hydrodynamical model,
rescattering of secondaries generates pressure and drives the subsequent collective motion. In a non-central collision, the reaction zone is asymmetric (almond shaped), and pressure gradient
is large in one direction and small in the other. The asymmetric pressure gradient generates the elliptic flow. As the fluid evolve and
expands, asymmetry in the reaction zone decreases and  a stage arise when reaction zone become symmetric and system no longer generate elliptic flow.  
Elliptic flow is an early time phenomena. It is a sensitive probe to, (i) degree of
thermalisation, (ii) transport coefficients and (iii) equation of state of the early stage of the fluid \cite{Ollitrault:1992bk,Kolb:2001qz,Hirano:2004en}. 
Relativistic hydrodynamics has been successfully applied to analyse various experimental data in Au+Au collisions   at RHIC energy ($\sqrt{s}$=200 GeV).  
Hydrodynamical evolution of near ideal QGP fluid, with shear viscosity to entropy ratio $\eta/s\approx$0.1, initialized to central energy density $\varepsilon_i\approx$ 30 $GeV/fm^3$ at initial time $\tau_i\approx$0.6 fm/c, is consistent with experimental elliptic flow in central and mid-central collisions \cite{QGP3,Luzum:2008cw,Chaudhuri:2008sj,Chaudhuri:2008ed,Chaudhuri:2009uk}.
Small viscosity is consistent with string theoretical calculation based on
ADS/CFT correspondence \cite{Kovtun:2004de} that viscosity of a strongly interacting matter is bounded by the KSS bound 
$\eta/s \geq 1/4\pi$. One wonders about the   viscosity to entropy ratio ($\eta/s$) at LHC energy. 
Present paradigm is that $\eta/s$ has a minimum, possibly with a cusp, around the critical temperature  $T=T_c$  \cite{Csernai:2006zz}. Since, at LHC, produced fluid will be at a higher temperature than the fluid at RHIC,  one expects higher value of $\eta/s$ at LHC. 
 
Recently, ALICE collaboration measured charged particles
elliptic flow in 10-50\% Pb+Pb collisions at LHC energy, $\sqrt{s}_{NN}$=2.76 TeV \cite{Aamodt:2010pa}.
Differential elliptic flow in 10-50\% collisions at LHC is also rather similar to that at RHIC. Integrated elliptic flow on the other hand increased by $\sim$30\%.
ALICE collaboration also measured charged particles
multiplicity \cite{Collaboration:2010cz}, $p_T$ spectra \cite{Aamodt:2010jd} in  Pb+Pb collisions. At LHC multiplicity increase by a factor of $\sim$2.2, though the centrality dependence is rather similar to that in $\sqrt{s}_{NN}$=200 GeV Au+Au collisions.  High $p_T$ suppression is also more at LHC energy.   In 0-5\% collisions, $p_T$ spectra of charged particle's are suppressed by a factor $R_{AA}\sim 0.14$ at $p_T$=6-7 GeV, which is smaller than at lower energies. In  peripheral collision, suppression is modest, $R_{AA}\approx$0.6-0.7.  

Increase of elliptic flow at LHC energy collisions has been predicted in
several hydrodynamical and hybrid models. In \cite{Luzum:2009sb,Hirano:2010jg,Drescher:2007uh} it was predicted that flow will increase by 20-30\%.   Some models also predicted decreased elliptic flow in LHC energy \cite{Krieg:2007sx,Molnar:2007an,Chaudhuri:2008je}. For example, in \cite{Chaudhuri:2008je}, one of the present author predicted that at $\sqrt{s}_{NN}$=5 TeV Pb+Pb collisions, elliptic flow will decrease by $\sim$15\%. The prediction was based on several assumptions, which are incorrect as of now. For example, contrary to the present understanding that multiplicity increases with energy by a power law, in \cite{Chaudhuri:2008je} a  logarithmic dependence was assumed. This led to much reduced multiplicity than observed experimentally.  Also, in \cite{Chaudhuri:2008je},  confinement-deconfinement transition was assumed to be 1st order, contrary to the present understanding that the transition is a cross-over. 
 
In the present paper, in a viscous hydrodynamic model, we have analysed the ALICE data for the centrality dependence of the charged particles multiplicity, $p_T$ spectra in 0-5\% collisions, centrality dependence of integrated and differential elliptic flow and extracted viscosity to entropy ratio for the fluid produced in $\sqrt{s}_{NN}$=2.76 TeV Pb+Pb collisions.
Analysis indicate that the ALICE data on the centrality dependence of charged particles multiplicity and 0-5\% $p_T$ spectra are best explained in ideal hydrodynamics. Elliptic flow data on the otherhand demand viscosity. 
Small viscosity $\eta/s=0.06\pm 0.02$ is required if all the data sets are analysed simultaneously.

The paper is organised as follows: in section \ref{sec2}, we briefly explain the second order Israel-Stewarts theory of dissipative hydrodynamics. 
 Hydrodynamics equations are closed with an equation of state (EoS).
In section \ref{sec3} we discuss the EoS used in the present study. Section \ref{sec4} deals with the initial conditions used for hydrodynamic simulation of $\sqrt{s}_{NN}$=2.76 TeV Pb+Pb collisions. Simulation results  are compared with experiments in section \ref{sec5}. Finally in section \ref{sec6} we summarise the results.

\section{Hydrodynamic model}\label{sec2}

In Israel-Stewart's second order theory \cite{IS79}, thermodynamic space is extended to include the dissipative fluxes.   Relaxation equations for the dissipative fluxes
are obtained from the entropy law, $\partial_\mu s^\mu \geq 0$. Equation of motion of the fluid is then obtained by solving simultaneously the conservation equations, e.g. energy-momentum and the relaxation equations for the dissipative fluxes. In $\sqrt{s}_{NN}$=2.76 TeV Pb+Pb collisions, the central rapidity region is approximately baryon free and dissipative effect like conductivity can be neglected. We also neglect the bulk viscosity. In general bulk viscosity is much less than the shear viscosity. But in QCD, near the transition point bulk viscosity can be large \cite{Kharzeev:2007wb,Karsch:2007jc}. Effect of bulk viscosity on elliptic was studied in   \cite{Song:2008si}. The effect is not large. We have also extended the code AZHYDRO-KOLKATA to include bulk viscosity. Preliminary results are reported in \cite{Chaudhuri:2011dp}. We also find that inclusion of bulk viscosity marginally modify the spectra or elliptic flow.
Equation of motion of the fluid in $\sqrt{s}_{NN}$=2.76 TeV Pb+Pb collisions is then obtained by solving,
  
\begin{eqnarray}  
\partial_\mu T^{\mu\nu} & = & 0,  \label{eq1} \\
D\pi^{\mu\nu} & = & -\frac{1}{\tau_\pi} (\pi^{\mu\nu}-2\eta \nabla^{<\mu} u^{\nu>}) \nonumber \\
&-&[u^\mu\pi^{\nu\lambda}+u^\nu\pi^{\mu\lambda}]Du_\lambda. \label{eq2}
\end{eqnarray}

Eq.\ref{eq1} is the conservation equation for the energy-momentum tensor, $T^{\mu\nu}=(\varepsilon+p)u^\mu u^\nu - pg^{\mu\nu}+\pi^{\mu\nu}$, 
$\varepsilon$, $p$ and $u$ being the energy density, pressure and fluid velocity respectively. $\pi^{\mu\nu}$ is the shear stress tensor (we have neglected bulk viscosity and heat conduction). Eq.\ref{eq2} is the relaxation equation for the shear stress tensor $\pi^{\mu\nu}$.   
In Eq.\ref{eq2}, $D=u^\mu \partial_\mu$ is the convective time derivative, $\nabla^{<\mu} u^{\nu>}= \frac{1}{2}(\nabla^\mu u^\nu + \nabla^\nu u^\mu)-\frac{1}{3}  
(\partial . u) (g^{\mu\nu}-u^\mu u^\nu)$ is a symmetric traceless tensor. $\eta$ is the shear viscosity and $\tau_\pi$ is the relaxation time.  It may be mentioned that in a conformally symmetric fluid relaxation equation can contain additional terms  \cite{Song:2008si}.

Assuming boost-invariance, Eqs.\ref{eq1} and \ref{eq2}  are solved in $(\tau=\sqrt{t^2-z^2},x,y,\eta_s=\frac{1}{2}\ln\frac{t+z}{t-z})$ coordinates, with the code 
  "`AZHYDRO-KOLKATA"', developed at the Cyclotron Centre, Kolkata.
 Details of the code can be found in \cite{Chaudhuri:2008sj}. 

\section{Equation of State} \label{sec3}

Hydrodynamic equations  are closed with an equation of state $p=p(\varepsilon)$.
Most reliable information about the QCD equation of state is obtained from lattice simulations. It is now established that the confinement-deconfinement transition is not a thermodynamic phase transition, rather a cross-over \cite{Aoki:2006we}. Since it is cross-over, there is no unambiguous temperature where the transition take place.   Inflection point of the Polyakov loop is generally quoted as the (pseudo)critical temperature for the confinement-deconfinement transition. Currently, there is consensus that
pseudo critical temperature for confinement-deconfinement phase transition  is
$T_c\approx$170 MeV \cite{Aoki:2009sc,Fodor:2010zz,Borsanyi:2010cj}. 
 
\begin{figure}[t]
\center
 \resizebox{0.35\textwidth}{!}{%
  \includegraphics{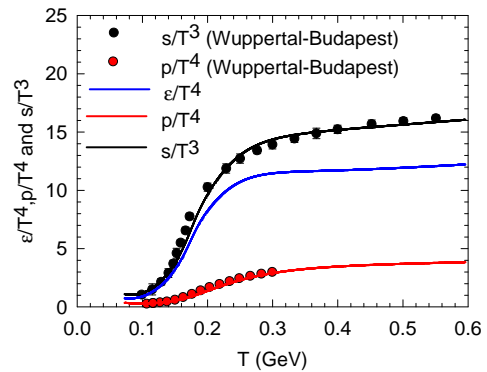}
}
\caption{(color online) Black and red circles are Wuppertal-Budapest lattice simulations \cite{Borsanyi:2010cj} for entropy density and pressure. Solid lines are the fit to the lattice simulation with model the EOS (Eq.\ref{eq3}). For completeness energy density in the model EOS is also shown.}
  \label{F1}
\end{figure} 
 
 We have  constructed EoS by complementing Wuppertal-Budapest lattice simulations \cite{Borsanyi:2010cj}  with a hadron resonances gas comprising all the resonances upto mass $m_{res}\approx$ 2.5 GeV. Entropy density is parameterised as,

\begin{equation} \label{eq3}
s=0.5[(1-\tanh(x)]s_{HRG}+0.5[1+\tanh(x)]s_{lattice}
\end{equation}

\noindent where $s_{HRG}$ and $s_{lattice}$  are entropy density of the confined and deconfined phase, $x=\frac{T-T_c}{\Delta T_c}$, $\Delta T_c=0.1 T_c$. 
From the parametric form of the entropy density, pressure and energy density can be obtained using the thermodynamic relations,

\begin{eqnarray} \label{eq4} 
 p(T)&=&\int_0^T s(T^\prime) dT^\prime \label{eq2a} \\
 \varepsilon(T)&=&Ts -p.
\end{eqnarray}

As stated earlier for $s_{HRG}$ we have used hadronic resonance gas comprising all the resonances with mass $m_{res}\leq$ 2.5 GeV.
For $s_{lattice}$ we use a parametric form for the  Wuppertal-Budapest simulations for entropy density,

\begin{equation}
\frac{s_{lattice}}{T^3}=[\alpha +\beta T][1+\tanh\frac{T-T_c}{\delta}],
\end{equation}

\noindent with $\alpha$=6.74, $\beta$=0.32, $T_c$=174 MeV and $\delta$=70 MeV.
 
In Fig.\ref{F1},  
Wuppertal-Budapest simulations for the entropy density and pressure, as  a function of temperature are shown. The solid lines are   fit obtained to the  lattice simulations for entropy density and pressure in the model EoS. The model EoS correctly reproduces lattice simulations for entropy and pressure.

\section{Initialisation of the fluid}\label{sec4}

Solution of partial differential equations (Eqs.\ref{eq1},\ref{eq2}) requires initial conditions, e.g.  transverse profile of the energy density ($\varepsilon(x,y)$), fluid velocity ($v_x(x,y),v_y(x,y)$) and shear stress tensor ($\pi^{\mu\nu}(x,y)$) at the initial time $\tau_i$. One also need to specify the viscosity ($\eta$) and the relaxation time ($\tau_\pi$). A freeze-out prescription is also needed to convert the information about fluid energy density and velocity to particle spectra and compare with experiments.
 
In a recent publication, we have analysed the centrality dependence of charged particles multiplicity in $\sqrt{s}_{NN}$=2.76 TeV Pb+Pb collisions \cite{Chaudhuri:2011hv}. It was shown that very short thermalisation time $\tau_i$=0.2 fm  or very large thermalisation time   $\tau_i$=4 fm, is not consistent with the ALICE measurements.  Presently, we assume that the fluid is thermalised at the time scale $\tau_i$=0.6 fm. Initial fluid velocity is assumed to be zero, $v_x(x,y)=v_y(x,y)=0$. At the initial time $\tau_i$, in an impact parameter ${\bf b}$ collisions, we distribute the initial energy density as in a Glauber model \cite{QGP3},

\begin{equation} \label{eq7}
\varepsilon({\bf b},x,y)=\varepsilon_i[(1-x) N_{part}({\bf b},x,y) + x N_{coll}({\bf b},x,y)],
\end{equation}
\noindent $\varepsilon_i$ in Eq.\ref{eq7} is the central energy density in impact parameter ${\bf b}=0$ collision. $N_{part}$ and $N_{coll}$ are the transverse profile of the average number of  participants and average number binary collisions respectively, calculated in a Glauber model. $x$ is the fraction of hard scattering and is an important parameter for elliptic flow development. Initial spatial eccentricity is more if initial density is dominated by the hard scattering, so is the elliptic flow \cite{Roy:2010zd}. In Glauber model of initialisation with energy density, unless the hard scattering fraction is large,
ALICE data on the centrality dependence of the charged particles multiplicity is not reproduced. We assume 90\% hard scattering fraction. 
The shear stress tensor was initialised to boost-invariant values, $\pi^{xx}=\pi^{yy}=2\eta/3\tau_i$, $\pi^{xy}$=0. For the relaxation time, we used the   Boltzmann estimate $\tau_\pi=3\eta/2p$. 
Finally, the freeze-out temperature was fixed at $T_F$=130 MeV.  Elliptic flow is an early time phenomena and depend little on the freeze-out temperature. 
In viscous hydrodynamics, particle production has contributions from the non-equilibrium part of the distribution function. We have
checked that at the freeze-out, non-equilibrium contribution to particle production is much smaller than the equilibrium contribution.
We note that the freeze-out condition does not account for the validity condition for viscous hydrodynamics, i.e. relaxation time for dissipative fluxes are much greater than the inverse of the expansion rate, $\tau_R \partial_\mu u^\mu << 1$. Recently, Dusling and Teaney \cite{Dusling:2007gi} implemented dynamical freeze-out condition. 
In dynamical freeze-out, non-equilibrium effects are stronger than obtained at fixed temperature freeze-out. 

\begin{figure}
\center
 \resizebox{0.35\textwidth}{!}{%
  \includegraphics{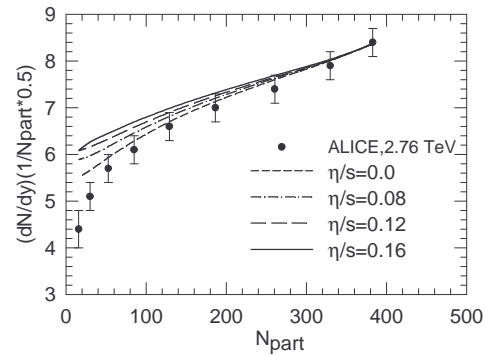}
}
\caption{ALICE measurements for centrality dependence of charged particles multiplicity are compared with hydrodynamical simulations for $\eta/s$=0-0.16.}
   \label{F2}
\end{figure}

\begin{table}[h]
\caption{\label{table1} Central energy density ($\varepsilon_i$) and 
temperature ($T_i$) at the initial time $\tau_i$=0.6 fm/c, for different 
values of viscosity to entropy ratio ($\eta/s$).   The bracketed values are estimated central energy density and temperature in $\sqrt{s}_{NN}$=200 GeV Au+Au collisions \cite{Chaudhuri:2009uk}. 
} 
\begin{ruledtabular} 
  \begin{tabular}{|c|c|c|c|c|}\hline
$\eta/s$         & 0    & 0.08 & 0.12 & 0.16 \\  \hline
$\varepsilon_i$ & $143.0\pm 6.0$ & $126.0\pm 5.5$ & $115.0\pm 5.2$ &  $103.0\pm 4.6$ \\    
 ($\frac{GeV}{fm^3}$)  & ($35.5\pm 5.0$) & ($29.1\pm3.6$) & ($25.6\pm 4.0$) &  ($20.8\pm 2.7$)  \\ \hline
$T_i$   & $548\pm 5$ & $531\pm 7$ & $520\pm 6$ & $504\pm 7$     \\
(MeV)  & ($377\pm 14$) & ($359\pm 12$) & ($348\pm 14$) & ($331\pm 11$) 
 \end{tabular}\end{ruledtabular}  
\end{table}  

The only parameters remain to be fixed is the central energy density $\varepsilon_i$ and viscosity coefficient $\eta$. We assume that the ratio of viscosity to entropy $\eta/s$ remain a constant through out the evolution and simulate Pb+Pb collisions for four values of viscosity, (i) $\eta/s$=0 (ideal fluid) (ii) $\eta/s=1/4\pi\approx 0.08$ (KSS bound), (iii)$\eta/s$=0.12 and (iv) $\eta/s$=0.16. During the evolution, fluid cross over from QGP phase to hadronic phase. The assumption that $\eta/s$ remain unchanged during the evolution violates the present understanding  that $\eta/s$ has a minimum  around the critical temperature  $T=T_c$  \cite{Csernai:2006zz}.   Constant $\eta/s$ should be understood as space-time averaged $\eta/s$. The remaining parameter, central energy density $\varepsilon_i$ at the initial time $\tau_i$=0.6 fm is fixed to reproduce ALICE measurements for charged particles multiplicity in 0-5\% Pb+Pb collisions, $(\frac{dN_{ch}}{dy})_{ex}=1601\pm 60$ \cite{Collaboration:2010cz}. We simulate Pb+Pb collisions and compute the
negative pion multiplicity. Resonance contributions are included. Assuming that pions constitue $\approx$85\% of the charged particles, negative pion multiplicity is multiplied by the factor $2\times 1.15$ to compare with experimental charged particles multiplicity. 
Irrespective of fluid viscosity, in the hydrodynamic model, experimental multiplicity can be fitted by changing the initial energy density, more viscous fluid requiring less energy density.   Multiplicity is proportional to final state entropy. In viscous fluid, during evolution, entropy is generated. Entropy generation increases with viscosity and more viscous fluid require less initial energy density to reach a fixed final state entropy. In table.\ref{table1}, central energy density and temperature of the fluid required to reproduce charged particles multiplicity in 0-5\% collisions are listed.   ALICE collaboration measured multiplicity quite accurately and initial energy density of QGP fluid in $\sqrt{s}_{NN}$=2.76 Pb+Pb collisions can be estimated accurately, within 5\%.   
In table.\ref{table1}, the bracketed values are estimated central energy density and temperature in $\sqrt{s}_{NN}$=200 GeV Au+Au collisions \cite{Chaudhuri:2009uk}.  From RHIC to LHC,  collision energy increases by a factor of $\sim$14 but  central energy density of the fluid increase by a much lower factor $\sim$ 4-5.

\begin{figure}
\center
 \resizebox{0.35\textwidth}{!}{%
  \includegraphics{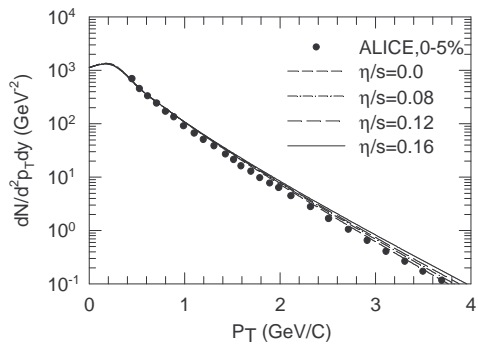}
}
\caption{ALICE measurements for charged particles $p_T$ distribution 
 0-5\% Pb+Pb collisions are compared with hydrodynamic simulations for $\eta/s$=0-0.16.} 
  \label{F3}
\end{figure}

\section{Results}\label{sec5}

\subsection{Centrality dependence of charged particles multiplicity}

With all the model parameters fixed we can study the centrality dependence of simulated charged particles multiplicity in  $\sqrt{s}_{NN}$=2.76 TeV Pb+Pb collisions. In Fig.\ref{F2} simulation results for charged particles multiplicity per participant pairs are compared with  the ALICE measurements. 
Even though, fluid was initialised to reproduce experimental multiplicity in 0-5\% collisions, in   peripheral collisions, viscous fluid produces more particles than an ideal fluid. 
 The reason is understood. Viscous effects depend on the velocity gradients. Velocity gradients are more in a peripheral collisions than in a central collisions, so is the viscous effect.  
To be quantitative about the fit to the data in ideal and viscous hydrodynamics, we have computed $\chi^2$ values for the fits, they are listed in table.\ref{table2}. Best fit to the data is obtained in the ideal fluid approximation.  In viscous evolution $\chi^2$ values increase by a factor of 2-3.5. We can conclude that ALICE charged particles multiplicity data in $\sqrt{s}_{NN}$=2.76 TeV Pb+Pb collisions do not demand any viscosity.

 \begin{figure}[t]
\center
 \resizebox{0.35\textwidth}{!}{%
  \includegraphics{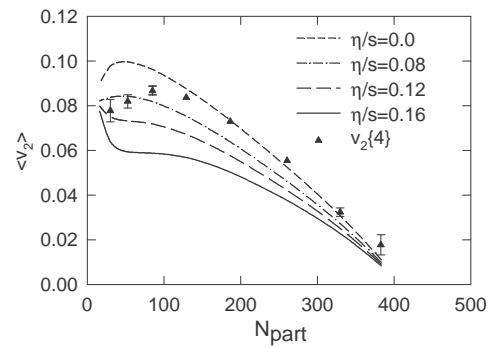}
}
\caption{ ALICE measurements for integrated elliptic flow are compared with hydrodynamic simulations for viscosity to entropy ratio $\eta/s$=0-0.16.}
   \label{F4}
\end{figure}

 \begin{table}[h] 
\caption{\label{table2} $\chi^2/N$ values for the fit to the ALICE data for different values of viscosity to entropy ratio.} 
  \begin{tabular}{|c|c|c|c|c|}\hline
  & \multicolumn{4}{|c|} {$\chi^2/N$}  \\ \hline \
Data set  & $\eta/s=0.0$ & $\eta/s$=0.08 & $\eta/s$=0.12 &$\eta/s$=0.16   \\ \hline 
\hline
charged particles & 1.46 & 3.12 & 4.44 & 5.24 \\
multiplicity & & & &\\ \hline
0-5\%    & 4.38 & 7.14 & 9.78 & 14.84 \\
$p_T$ spectra & & & &\\ \hline
$v_2$ integrated  & 9.68 & 3.02 & 12.46 & 41.44 \\ \hline
$v_2$: 10-20\%  & 7.07 &4.69  &7.68 &15.14  \\ \hline
$v_2$: 20-30\%  & 23.64&8.32  &8.46 &18.97  \\ \hline
$v_2$: 30-40\%  &44.60& 13.14  &9.89 &24.47   \\ \hline
$v_2$: 40-50\% & 33.62 &9.86 &4.18 &9.00    \\ \hline
\end{tabular}
 \end{table} 
 
\subsection{Charged particles $p_T$ spectra in 0-5\% collision}

ALICE measurements for charged particles $p_T$ spectra in 0-5\% collisions are compared with hydrodynamical simulations in Fig.\ref{F3}. 
We have initialised the fluid to reproduce charged particles multiplicity in 0-5\% collision. Such a initialisation do reproduces experimental $p_T$ spectra reasonably well. Effect of viscosity on spectra is also evident. Compared to ideal fluid, in viscous evolution, high $p_T$ production is enhanced.  In table.\ref{table2}, $\chi^2$ values for the fits are listed.
Here also, we can conclude that ALICE data on charged particles $p_T$ spectra in 0-5\% collisions, do not demand any viscosity.

\begin{figure}
\center
 \resizebox{0.4\textwidth}{!}{%
  \includegraphics{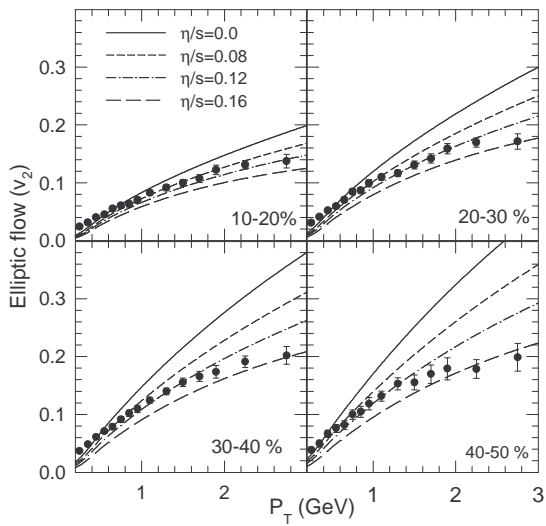}
}
\caption{In four panels, ALICE measurements for charged particles 
 elliptic flow in 10-20\%, 20-30\%, 30-40\% and 40-50\% Pb+Pb collisions 
 are shown. The solid, short dashed, dashed dotted and long dashed lines are hydrodynamic model simulation for elliptic flow for fluid viscosity $\eta/s$=0, 0.08, 0.12 and 0.16 respectively.}
  \label{F5}
\end{figure}

\subsection{Centrality dependence of integrated elliptic flow}
 
  ALICE collaboration employed various methods, e.g. 2 and 4 particle cumulant, q-distribution, Lee-Yang zero method etc, to measure charged particles elliptic flow in Pb+Pb  collisions \cite{Aamodt:2010pa}. Flow in 4-particle cumulant method, Lee-Yang zero method and q-distribution is consistent with each other. 2-particle cumulant method however results in $\sim$15\% higher flow. In Fig.\ref{F4} ALICE measurements for integrated elliptic flow in 4-particle cumulant method are compared with hydrodynamical simulations for elliptic flow.  Ellitic flow depend   sensitively on viscosity, flow reducing with increasing viscosity. Reductiuon is more in peripheral than in central collisions. As argued previously, viscous effects are more in peripheral than in central collisions. For high viscosity, $\eta/s$=0.16, in very peripheral collisions, experimental centrality dependence is not reproduced. Instead of decreasing $v_2$ increases.  $\chi^2/N$ values for the fits to the data are noted in table.\ref{table2}. Unlike the centrality dependence of charged particles multiplicity or $p_T$ spectra, best fit to the ALICE data on integrated elliptic flow is obtained for  $\eta/s$=0.08.
 
\subsection{Differential elliptic flow in 10-50\% collision}  

In Fig.\ref{F5}, ALICE measurements of elliptic flow $v_2\{4\}$, in 4-particle cumulant method, in   10-20\% ,20-30\%, 30-40\% and 
40-50\% collision centralities    are shown.  ALICE collaboration measured elliptic flow upto $p_T\approx$5 GeV. Differential $v_2$ reaches a maximum $\approx$0.2 at $p_T$=3 GeV. In Fig.\ref{F5}, we have shown the measurements only upto $p_T$=3 GeV. Hydrodynamical model are not well suited for large $p_T$. Sources other than thermal also contribute to the particle production at large $p_T$. The continuous, small dashed, dashed dot, and medium dashed lines in Fig.\ref{F5} are simulated elliptic flow for pions in hydrodynamic evolution of fluid with viscosity   to entropy density ratio $\eta/s$=0.0,0.08,0.12 and 0.16 respectively. 
 
\begin{figure}[t]
\center
 \resizebox{0.3\textwidth}{!}{%
  \includegraphics{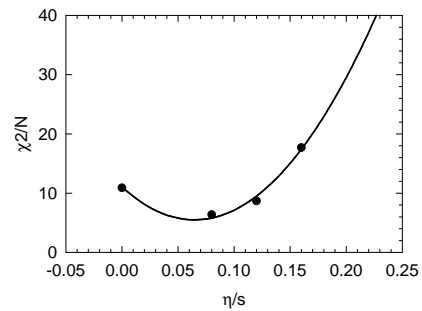}
}
\caption{ $\chi^{2}/N$ values for the fits to the ALICE combined data set (see text). The solid line is a parabolic fit to the $\chi^2$ values.  }
  \label{F6}
\end{figure}

Unlike the centrality dependence of multiplicity or the $p_T$ spectra in 0-5\% collisions, differential elliptic flow is not best explained in ideal fluid evolution.
$\chi^2/N$ values for the fits, in each collision centralities, are noted in table.\ref{table2}. Experimental data only in the $p_T$ range, $0.25 \leq p_T \leq 3.25$ GeV   are included in the $\chi^2$ analysis. $\chi^2/N$ values indicate that the  data are better explained in viscous evolution than in ideal fluid evolution. It is also apparent that more peripheral collisions demand more viscosity. For example, in 10-20\%
  collision,    $\chi^2$ is minimised for the KSS bound of viscosity to entropy ratio,  $\eta/s \approx$0.08.   In more peripheral 30-40\% collision, data demand   $\eta/s\approx$0.12. The result is interesting. In RHIC energy collisions also, it was seen that elliptic flow in peripheral collisions demand higher viscosity than in central collisions  \cite{Chaudhuri:2009hj}. As discussed in detail in \cite{Chaudhuri:2009hj}, centrality dependence of viscosity can be understood in terms of continuously increasing role of hadronic phase of the fluid in peripheral collisions. The space-time averaged $\eta/s$ has contributions, both from the QGP and hadronic phase. Compared to a central collisons, in a  peripehral collision, contribution of hadronic phase to the space-time averaged $\eta/s$ increases, while that of QGP decreases.

In the present analysis different $\eta/s$ are indicated by different data sets.
For example, ALICE data on multiplicity and 0-5\% $p_T$ spectra are best explained with $\eta/s\approx$0, the  integrated $v_2$ data with $\eta/s\approx$0.08, the differential $v_2$ data in 10-50\% collision with $\eta/s$=0.08-0.12. To obtain $\eta/s$ which best explain the ALICE data, 
    $\chi^2/N$ values for the combined data set, (i) centrality dependence of multiplicity, (ii) centrality dependence of integrated elliptic flow, (iii) $p_T$ spectra in 0-5\% collisions and (iv) differential elliptic flow in 10-50\% collisions, are computed. They are shown in Fig.\ref{F6}. $\chi^2/N$ values exhibit a broad minima, indicating that the data are not very sensitive to viscosity to entropy ratio. The solid line in Fig.\ref{F6} is a parabolic fit to the $\chi^2/N$ values. From the minimum of the fit we extract, $\eta/s=0.06 \pm 0.02$. As it was at RHIC enery collisions, at LHC also, experimental data are consistent with nearly ideal fluid. 
We may note that the extracted value of $\eta/s$ at LHC energy is compare well with $\eta/s$ extracted in other hydrodynamic model simulation 
 \cite{Schenke:2011tv,Bozek:2011wa}. However, hybrid models, i.e. hydrodynamics coupled to hadron cascade models indicate higher viscosity \cite{Song:2011qa,Hirano:2010je}.
 
 In the present analysis we have used some specific initial conditions, e.g. initial time $\tau_i$=0.6 fm, initial zero fluid velocity, hard scattering fraction 90\%, boost-invariant shear stress tensor etc. All possible initial conditions are not explored. In \cite{Chaudhuri:2009uk}, systematic uncertainty in $\eta/s$ due to various uncertainties in initial conditions was estimated as large as 175\%. The presently extracted value $\eta/s=0.06\pm 0.02$ will be even more uncertain, if all possible initial conditions are accounted for.

\section{Summary and Conclusion}\label{sec6}

In Israel-Stewart's second order theory of hydrodynamics, we have 
analysed ALICE data on the centrality dependence of charged particles  multiplicity, integrated and differential elliptic flow, charged particles $p_T$ spectra in 0-5\% collisions.  ALICE data for charged paricles multiplicity or 0-5\% $p_T$ spectra do not demand any viscosity, data are best explained in ideal fluid approximation. On the other hand, ALICE data on integrated or differential elliptic flow  demand small viscosity. From a simulataneous analysis of all the data sets we obtain viscosity to entropy ratio $\eta/s=0.06\pm 0.02$, in $\sqrt{s}_{NN}$=2.76 TeV Pb+Pb collisions. 
The extracted values of viscosity is rather similar to the value extracted at RHIC energy,  even though the fluid at LHC is at   higher temperature than at RHIC. It appear that both at RHIC and LHC,     nearly perfect fluid is produced. 

$\bf{Acknowledgement}$: Authors would like to thank Dr. Bedangadas Mohanty for providing the 
ALICE data on charged particle elliptic flow for different centralities.

\end{document}